# Direct Method for Calculating the Inverse Radon Transform and Its Applications

E.E. Libin[1], S.V. Chakhlov[2], and D. Trinca[2]

[1]National Research Tomsk State University, Tomsk, Russian Federation

[2]Laboratory of Technical Tomography and Introscopy, National Research Tomsk Polytechnic University, Tomsk, Russian Federation

Corresponding author: chakhlov@tpu.ru

## Contents

## Introduction

The purpose of this report is a study of the algebraic approach possibilities to reconstruct images. This approach is reduced to solution of the large system of linear algebraic equations.

## The main functional equation

In case of continuous distribution of phantom density the mathematic formula describing the ray sums in sinogram (for parallel projection scanning) has form

$$S(p, \varphi) = \iint \mu(x, y) \delta(p - x \cos \varphi - y \sin \varphi) \, dx \, dy \,,$$
(1)

where: $\mu(x, y)$ is phantom density, $\delta(p)$ is a delta-function of Dirac.

If the model phantom μ(x, y) is known, the formula (1) allows to calculate sinogram S(p, φ). The inverse task means that to find μ(x, y) when the function S(p, φ) is known.

There is exist an exact solution of equation (1), which was found by Radon himself (1917). It names inverse Radon transform and has form

$$\mu(x.y) = -\frac{1}{2\pi^2} \int\limits_0^\pi d\varphi \int \frac{\partial S(p, \varphi)}{\partial p} \frac{dp}{p - x \cos \varphi - y \sin \varphi}.$$
(2)

Integral by variable $p$ in formula (2) understanding in sense of main value and names the convolution.

## Matrix D of the system of linear algebraic equations

The functional equation (1), considered in discrete form on a set of integer numbers, has form (3)

$$S(i, j) = \sum_n \sum_m D(i, j; n, m) \mu(n, m) \,,$$
(3)

$$D(i, j; n, m) = \delta \left[ i - round \left( n \cos \varphi_i + m \sin \varphi_j \right) \right].$$
(4)

δ is function in (4) understanding as δ(0)=1, δ(≠0)=0. n, m are numbers of phantom coordinates x and y; i is detector number, j is number of rotation angle.

The phantom and sinogram in functional equation (3) are two dimensional arrays of data. But one can always convert those arrays to one-dimensional vector-columns. The matrix D also can be converted from four dimensional array to two-dimensional one. After that the image reconstruction task become a usual task of linear algebra, but with large dimensions of its values.

The conversion of two dimensional arrays to one dimensional ones and backwards are produced by command reshape().

Let, for example, integer numbers nx, ny, np, nf – means:

nx, ny are the number of phantom elements in horizontal and vertical directions,

np is a number of detectors in array,

nf is a number of angle of phantom rotations during scanning.

Then we can convert two dimensional arrays S and μ to vector-columns by commands:

    S = reshape (S, np*nf, 1),          Mu = reshape (Mu, nx*ny, 1).

The functional equation (3) is converted to the usual linear algebraic relation (5)

$$\vec{S} = D\vec{\mu}.$$
(5)

Usually the sonogram has more number of elements than phantom, therefore the matrix D would be rectangular one. It has much more lines than columns.

The properties of matrix D:

1. It has larger dimension size(D) = [np*nf, nx*ny].
2. It is very sparse (has very many zeros).
3. It consists from zeros and units only.

Code to calculate matrix D

```
function D=dtom2(np,nf,nx,ny)
P=linspace(-1.5,1.5,np); f=linspace(0,pi,nf);
 x=linspace(-1,1,nx);
Y=linspace(-1,1,ny); [F,X,Y]=ndgrid(f,x,y);
D=X.*cos(F)+Y.*sin(F);
D=(D+1.5)/3; D=round(np*D);
D=reshape(D, 1, nf*nx*ny);
D=repmat(D, np, 1) ==repmat([1:np]', nf*nx*ny);
D=reshape(D, np*nf, nx*ny); D=sparse(D);
```

To calculate matrix D by formula (4) there was used the abovementioned code. In code at first the number of detector is defined as function of x, y, φ. Than the matrix D was added to fourth dimension and reshaped to two dimension array.

After execution of command D = dtom2(100,50, 40,40) and visualization of matrix D (shown on Fig. 1 by command spy(D)), we get that the matrix D contains 8 millions elements, from which 80 000 are non-zero only.

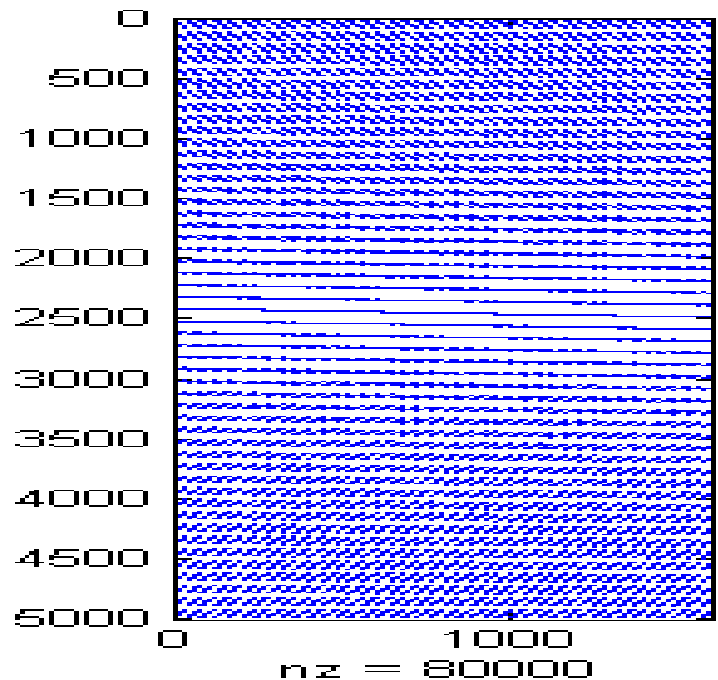

Fig. 1

Another example of matrix D visualization with use of function dtom2 is shown in Fig. 2.The sparse matrices are shown by command Spy(D).

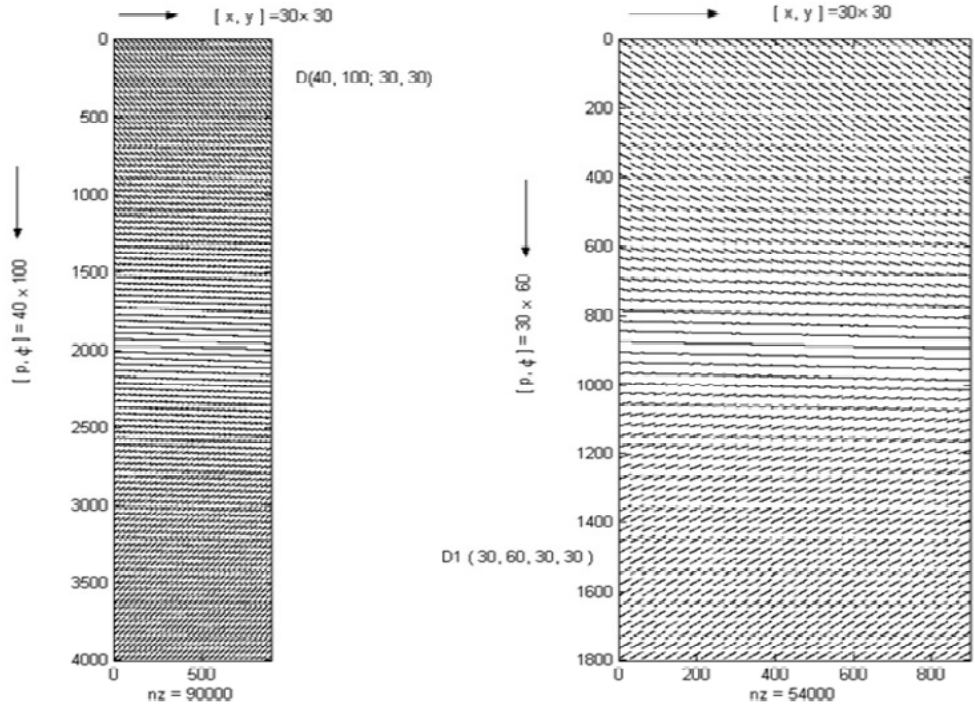

Fig. 2 Overview of sparse matrices D(40*100, 30*30) and D1(30*60, 30*30) with different number lines, but with the same number of columns.

## Solution of redefined system (5) by the least square method

The least square method is reduced finally to, that equation (5) must be multiplied from left on transposed matrix D'.

$$(D' * D)\mu = D' * S \quad (6)$$

The matrix (D'*D) is square one, with sizes nx*ny, and non-singular, therefore the equation (6) can solved by operation of backslash \.

$$\mu = (D'*D) \backslash (D'*S) \quad (7)$$

## Scenario of calculation with the model phantom

In this case the phantom was defined as two small rectangles with different densities. After execution of below mentioned code there is produced the original phantom, its sonogram and reconstruction result.

```
np=60; nf=180; nx=50; ny=50;
D=dtom2(np, nf, nx, ny); % spy(D);
Mu=zeros(nx, ny); Mu(8:12, 8:12)=1; Mu(20:25, 15:20)=0.5;
Figure; bar3(Mu); colormap(white);
Mu1=reshape(Mu, nx*ny, 1); S=D*Mu1;
figure; bar3(reshape(S, np, nf)); colormap(white);
Mu1=full(D'*D)\ (D'*S); Mu1=reshape(Mu1, nx, ny);
Figure; bar3(Mu1); colormap(white);
```

**Result of model scenario**

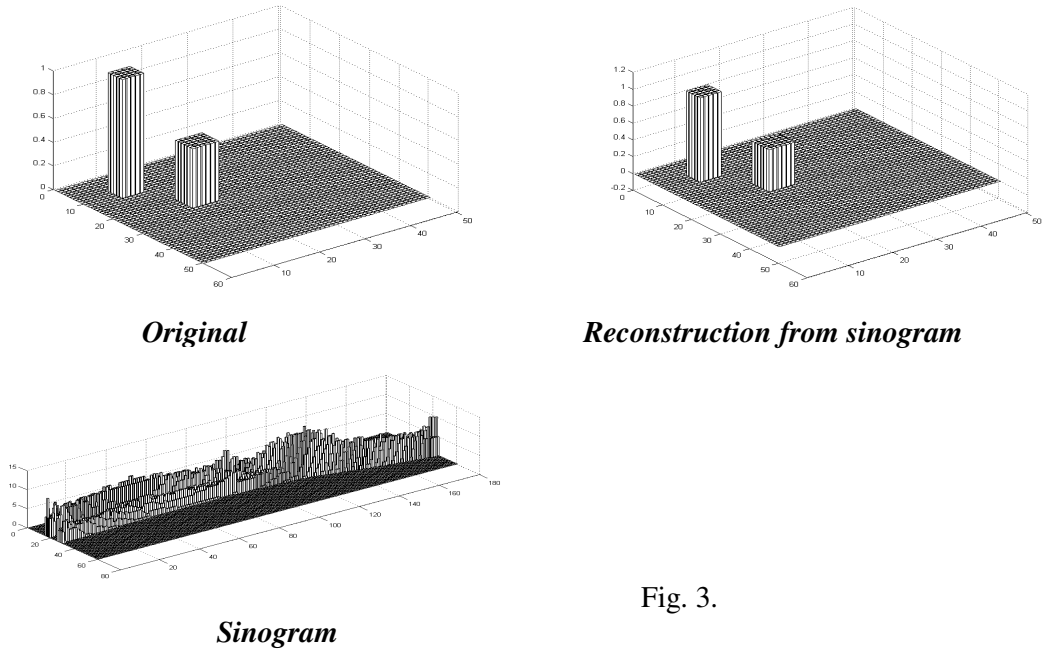

*Original*                    *Reconstruction from sinogram*

Fig. 3.

*Sinogram*

The result output as column diagrams shows almost ideal reconstruction of the original phantom.
In this case we solve a system of linear algebraic equations (7) with 2500 unknowns densities μ(x,y)
by the direct method.

## Comparison with approximated methods (iterations)

Exist opinion that too large systems of linear algebraic equations have to be solved by iterative
methods. Shown on Fig. 4 example of equation (7) solution by use the function qmr illustrates that
iterative method is working worse than the direct one.

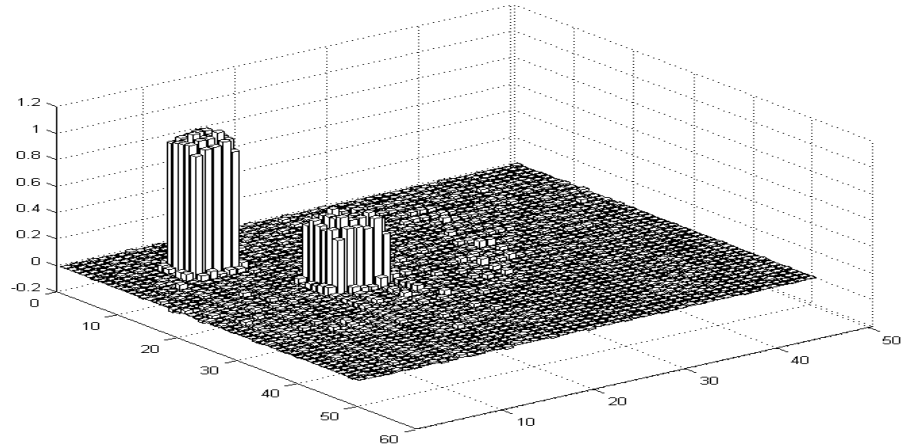

**Fig. 4**. Result of image reconstruction by iterative method with help of command
Mu1 = qmr(D'*D,  D'*S, 1e-3, 50).

## Comparison with classic method of Radon transform

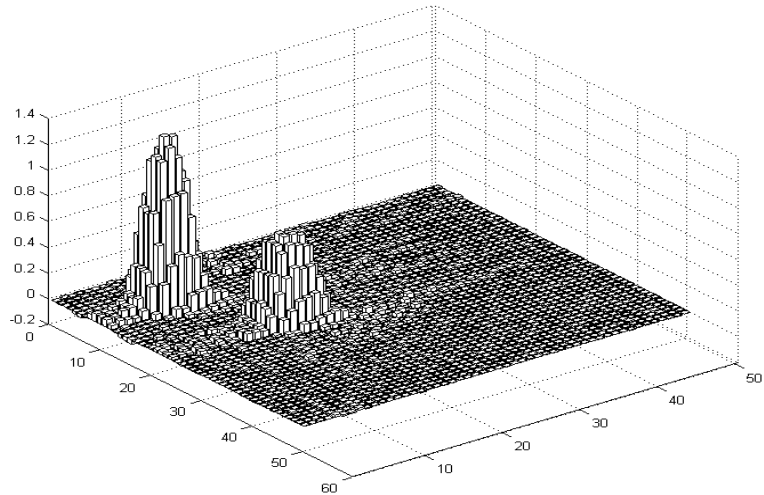

**Fig. 5**. Reconstruction of model phantom by Radon transform.

Differentiation and convolution in the Radon transform (2) there were calculated approximately in the matrix form. We can see that this approach is working, but the result is worse than in direct algebraic method. If the matrix $D$ of system of the linear algebraic equations is stored in computer memory then it is better to solve it by direct method.

## Scanning by fan-beam projections

Let (Fig. 6) point $M$ of fixed phantom has coordinates $x+iy$. The radiation source is placed in point $R$ and is visible under angle $\varphi$ to axis $Ox$. Then, evidently, we can write:

$$x+iy = se^{i\varphi} + h\, i\, e^{i\varphi}$$

or

$$s+ih = e^{-i\varphi}(x+iy)$$

The angle $\alpha$ (under which we can see from radiation source $R$ the phantom point $M$ and axis $OR$) is equal $arctg\ h/(R\text{-}s)$, or:

$$\alpha = angle\left(R - e^{-i\varphi}(x+iy)\right) \tag{8}$$

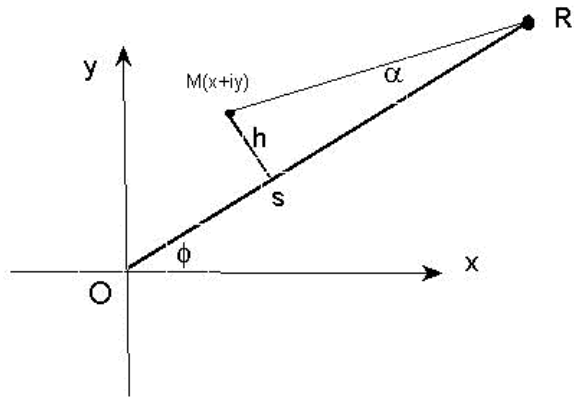

**Fig. 6**

A set of radiation sources with parameters $R_k$, $\Phi_k$ are defined different projections on the detector array. The angle $\alpha$, calculated by formula (8), is defined a number of detector, on which falling a ray passing through the point *M*.

If the distance from radiation sources to phantom center is fixed, then calculation of the main tomographic matrix *D* by formula (8) is produced by function *dtom3()*. It takes into account the view angle of detectors (aperture). In this code it is equals 90 degrees.

```
function D=dtom3(np,nf,nx,ny)
Ap=90*pi/180; R=1.45; f=linspace(pi/2,3*pi/2,nf);
x=linspace(-1,1,nx); y=linspace(-1,1,ny);
[F,X,Y]=ndgrid(f,x,y); D=angle(R-exp(-i*F).*(X+i*Y));
D=(D+Ap/2)/Ap; D=round(np*D);
D=reshape(D,1,nf*nx*ny); D=repmat(D,np,1)==repmat([1:np]',1,nf*nx*ny);
D=reshape(D,np*nf,nx*ny);D=sparse(D);
```

The model calculation was done by below mentioned scenario-file with name (testtom3).

```
np=68; nf=30; nx=40; ny=40;      %testtom3
D=dtom3( np,nf,nx,ny);
Mu=zeros(nx,ny); Mu(8:12,8:12)=1; Mu(20:21,15:20)=0.5; Mu(27:37,18:30)= 0.3;
Mu(29:35,21:28)=0;
figure; imagesc(max(max(Mu))-Mu); colormap(gray);
Mu1=reshape(Mu,nx*ny,1); S=D*Mu1;
S1=reshape(S,np,nf); figure; imagesc(max(max(S1))-S1); colormap(gray);
Mu1=full(D.'*D)\(D.'*S); Mu1=reshape(Mu1,nx,ny);
 figure; imagesc(max(max(Mu1))-Mu1); colormap(gray);
```

np is a number detectors in array, nf is a number of radiation sources, nx, ny are the phantom dimensions in horizontal and vertical.

The result of execution of scenario testtom3 is shown on Fig.7–10. There were used the following parameters: np=60, nf =32, nx=ny=43.

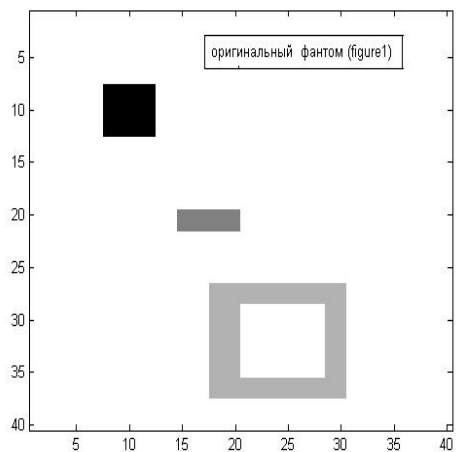

**Fig. 7.**

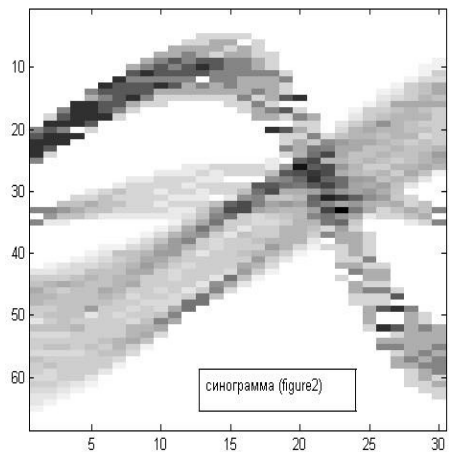

**Fig. 8**

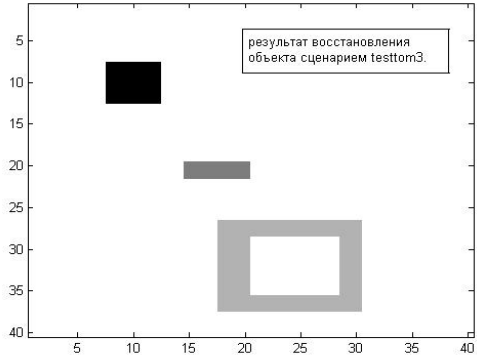
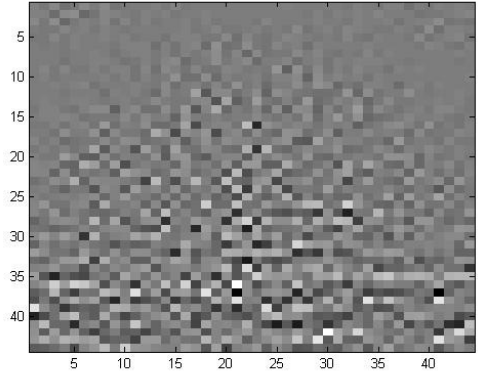

**Fig. 9**.　　　　　　　　　　　**Fig. 10**.

Good reconstruction of model phantom is due to that volume of sinogram 60*32= 1920 is exceed the volume of phantom 43*43=1849. However, if we take nx=ny=44, then the phantom volume is increased to value 44*44=1936 and would be larger then sonogram volume. In this case the object is not reconstructed (Fig. 10) and generated warning about singularity of matrix $D'*D$. Thus the accuracy of image is limited by sonogram volume. There always must be inequation np*nf>nx*ny.

## Connection of matrix method with Radon transform

Algebraic method of image reconstruction, based on matrix $D$ usage and the method of Radon transform, both are based on the same functional relation of type (1), which depend from the kind of radiation projecting on detector array.

If we multiply matrix $D$ from left on vector of phantom $\mu_0$, then we get a sinogram $S$.

If we multiply the transposed matrix $D'$ from left on vector of sinogram $S$, then there would be a summation of ray sums for all rays, going through the point with coordinates $(x,y)$. This representation of density is named the backprojection.

If, for example, add the following commands to scenario-file testtom3:

        OP=D'*S; OP=reshape(OP,nx,ny); figure;
        imagesc(max(max(OP))-OP); colormap(gray);

then we get a field of backprojections, which is shown on Fig. 11.

In Radon transform the operation of backprojection (that means left multiplication on matrix $D'$) is used not for the original sinogram $S$, but for the another sinogram $S_1$, which produced from $S$ by operations of differentiation and convolution for the detector numbers.
.

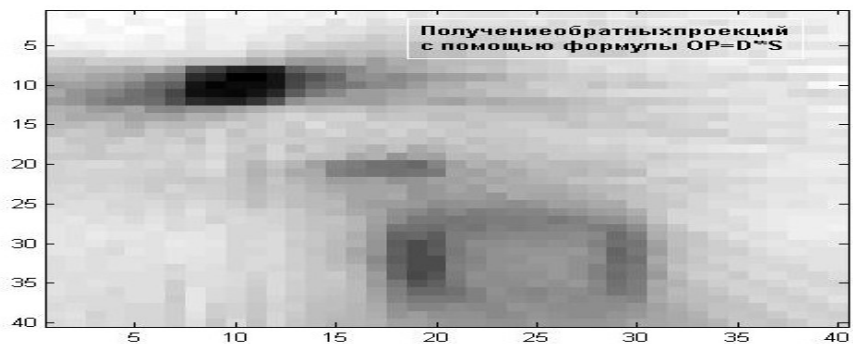

Fig. 11.

**Integral equation**

In theory of Radon transform there is interesting relation, which gives an integral equation to find the phantom density. Let consider double integral on a plane:

$$J(q,x,y) = \int\limits_{(x-\alpha)^2+(y-\beta)^2>q^2} \frac{\mu(\alpha,\beta)d\alpha d\beta}{\sqrt{(x-\alpha)^2+(y-\beta)^2-q^2}}. \tag{9}$$

This integral is calculated for all plane without circle with radius $q$ and with center in point $(x,y)$. If in integral change Cartesian coordinates $(\alpha,\beta)$ to ray coordinates $(s,\varphi)$, whose are related by expressions:

$$\alpha = x - s\sin\varphi + q\cos\varphi, \qquad \beta = y - s\cos\varphi + q\sin\varphi. \tag{10}$$

Then we can get

$$J(q,x,y) = \int\limits_0^{2\pi} d\varphi \int\limits_{s>0} \mu\big(x-s\sin\varphi+q\cos\varphi,\ \ y+s\cos\varphi+q\sin\varphi\big)\ ds. \tag{11}$$

The meaning of formula (11) is that value $J(q,x,y)$ is a sum of all ray sums for rays whose are tangential to circle with center in point $(x,y)$ and radius $q$.

If in formula (9) or (11) direct $q$ to zero then we get

$$\iint \frac{\mu(\alpha,\beta)d\alpha d\beta}{\sqrt{(x-\alpha)^2+(y-\beta)^2}} = G(x,y), \tag{12}$$

where $G(x,y) = J(0,x,y)$ is a sum of all ray sums for rays passing through the point $(x,y)$. That means $G(x,y)$ is exactly that result which produced by application of back projection method to the primary sinogram.

From other side the application of algebraic solution with usage the tomographic matrix $D$, leads to a system of linear equations in form (6), or (13).

$$\big(D^{'}*D\big)\mu = D^{'}S = G. \tag{13}$$

In equations (12) and (13) the right sides are equal, but in the right sides there are different operators: in left side of equation (13) there is square matrix $D' * D$, but in equation (12) there is an integral operator.

All attempts to solve integrak equation (12), by discretization of kernel and its conversion to the matrix operator, usually finished by failure. So, for the direct solution of equation (12) there is a warning about very ill-conditioned matrix, but for the iterative method there is a picture, similar to Fig. (11), i.e. there is a few distinctions from right side of G.

The merit of equation (13), based on application of the matrix $D$ only, is that there is no computational problems, and the model phantom is always reconstructed precisely and without any artefacts.

Fig. 12 shows the fillness degree of square matrix $D'*D$, which was generated by commands $D = dtom3(60,32, 40,40);$ spy($D'*D$).

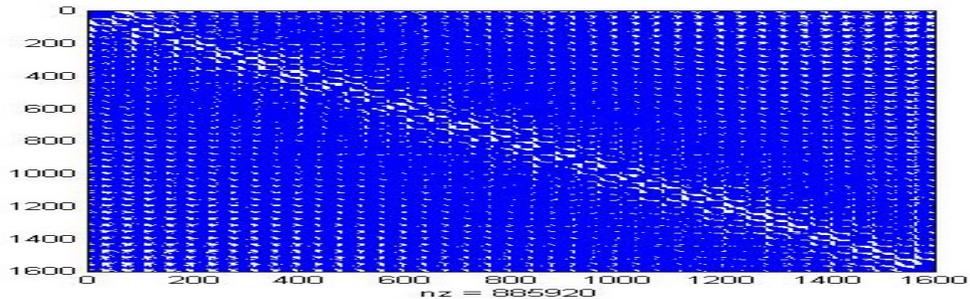

**Fig. 12**. The fillness degree of square matrix $D^{'}*D$.

Thus, in the matrix of the system equations (13) there is 885920 non-zero elements only, but at the same time the all matrix contains 2 560 000 elements. That means the fillness degree is about 35%.

Those matrices are enough sparse to be effectively processed. All operations with sparse matrices are supported in MatLab, and it is one additional argument to application system of equation (13) to reconstruct images. In the algebraic method there is used the matrix $D$ only and its connected square matrix $D'*D$, whose both are sparse and can be stored compactly. To apply the algebraic method it is necessary only that the sonogram volume was large then the phantom volume.

## Matrix interpretation of Radon transform

The method of Radon transform is widely used in tomography. Its main advantage is that there is no need to solve the system of algebraic equations (everybody want avoid that). This method gives the explicit expression for the phantom's density $\mu(x,y)$ through the ray-sums of sinogram $S(p,\varphi)$, Of course the matter is not reduced to find the inverse operator for equation (12), which is incorrect. In any case we can get the corresponding expression by using the different mathematical transformations and hypotheses. It seems that those transformations must be examined in detail and be replaced by its approximated matrix analogs. If in integral (9) change variables to:

$$\alpha = x + r\cos\psi, \quad \beta = y + r\sin\psi,$$ (14)

Then we get equation:

$$\int_{r=q}^{\infty} \frac{rdr}{\sqrt{r^2-q^2}} \int_0^{2\pi} \mu\left(x+r\cos\psi,\ y+r\sin\psi\right)\ d\psi = J(q,x,y).$$ (15)

The internal integral in (15) is a mean density along circles with radius $r > q$, that means

$$\hat{\mu}(r,x,y) = \frac{1}{2\pi}\int_0^{2\pi} \mu\left(x+r\cos\psi,\ y+r\sin\psi\right)\ d\psi$$

Thus, expression (15) is an integral equation to define unknown function $\mu^\wedge(r,x,y)$:

$$\int_{r=q}^{\infty} \frac{\hat{\mu}(r,x,y)rdr}{\sqrt{r^2-q^2}} = \frac{J(q,x,y)}{2\pi}.$$ (16)

Equation (16) is a famous Abel equation, and its solution is expressed by formula

$$\hat{\mu}(r,x,y) = -\frac{1}{\pi^2}\int_{q=r}^{\infty} \frac{\partial J/\partial q}{\sqrt{q^2-r^2}}\ dq$$ (17).

For $r \to 0$ the mean density naturally converted to the density in point *(x,y)*, and we get:

$$\mu(x,y) = -\frac{1}{\pi^2}\int_0^{\infty} \frac{\partial J(q,x,y)}{\partial q}\frac{dq}{q}.$$ (18)

Formula (18) is known as inverse Radon transform, to solve a task about reconstruction phantom's density by the measured ray sums.

If we know relation between distances q and detector numbers *p*, then the function *J(q,x,y)* can be written through the sinogram in form

$$J(q,x,y) = \int_0^{2\pi} S(p,\varphi)\ d\varphi.$$

Hence, the phantom density (18) can be presented as (19);

$$\mu(x,y) = -\frac{1}{\pi^2}\int_0^{2\pi} d\varphi \int \frac{\partial S(p,\varphi)}{\partial p}\frac{dp}{q\left(p,\varphi,x,y\right)}.$$ (19)

The value $q$, included into (19) is a distance from each phantom point $(x,y)$ to each ray $(p,\varphi)$ passing through detectors. The elements of four dimensional array $q$ have to be calculated by formula, which depend from the kind of projecting.

If the sonogram was created by parallel X-raying then dependence between $p$ and $q$ has form: $q = p - x \cos\varphi + y \sin\varphi$; and formula (19) is converted to:

$$\mu(x,y) = -\frac{1}{\pi^2} \int\limits_0^{2\pi} d\varphi \int \frac{\partial S(p,\varphi)}{\partial p} \frac{dp}{p - x\cos\varphi + y\sin\varphi} . \tag{20}$$

Formula (20) is used to demonstrate the Radon transform. Usually it gives a good reconstruction results for a large data volume in sonogram.

For another kinds of scanning the denominator $q(p,\varphi,x,y)$ in formula (19) must be replaced by other corresponding expressions.

For the fixed phantom the light sources, placed at distance $R$ from center are characterized by angles $\varphi$ (see Fig. 6). Different angles $\alpha$ are correspond to different detectors The ray direction is defined by ort-vector $L = e^{i(\varphi-\alpha)}$, but a distance from ray to the coordinate frame is equal $d = R \sin \alpha$. The normal to the ray is defined by vector $n = iL$, i.e. $n_x = -sin(\varphi-\alpha)$, $n_y = cos(\varphi-\alpha)$. Therefore the distance $q$ from point $(x,y)$ to ray is defined by formula.

$$q(\alpha,\varphi,x,y) = -x\sin(\varphi-\alpha) + y\cos(\varphi-\alpha) - R\sin\alpha . \tag{21}$$

Angles $\alpha$ are connected with detector numbers $p$ by dependence:

$$\alpha = \frac{A}{2}\left(1 - 2\frac{p-1}{n_p-1}\right). \tag{22}$$

Here $n_p$ is number of detectors in the group which is illuminated by source.

First detector ($p=1$) gives by formula (22) the angle $\alpha = A/2$, but last detector (for $p=n_p$) gives angle $\alpha = -A/2$. Angle $A$ is aperture of source, i.e. the angle under which from source one can see all detector's group.

By formulae (21) and (22) there are calculated the distances q including sign, i.e. for points placed above ray, the distances $q$ are positive, but in other half they has negative values. It is agree with formula (20). For the points on ray itself $q=0$.

In case of fan-beam scanning the formula (19) has from

$$\mu(x,y) = -\frac{1}{\pi^2} \int\limits_0^{2\pi} d\varphi \int \frac{\partial S(\alpha,\varphi)}{\partial \alpha} \frac{d\alpha}{-x\sin(\varphi-\alpha) + y\cos(\varphi-\alpha) - R\sin\alpha} . \tag{23}$$

For its practical usage one can introduce four dimensional array of inversed distances $Q(\alpha,\varphi;x,y)$, whose elements are calculated, for example, by formula.

$$Q(\alpha,\varphi,x,y) = -\frac{1}{\pi^2}\frac{q(\alpha,\varphi,x,y)}{q^2(\alpha,\varphi,x,y) + 0.001}. \tag{24}$$

As result we can write

$$\mu(x,y) = \int\limits_0^{2\pi} d\varphi \int\limits_\alpha Q(\alpha,\varphi,x,y)\, S_1(\alpha,\varphi)\, d\alpha , \tag{25}$$

where $S1(\alpha,\varphi) = \partial S(\alpha,\varphi)/\partial\alpha$ is a derivative from sonogram by detectors.

If the four dimensional array Q renumerate to two dimensional matrix, and sonogram $S1$ — to one dimensional vector, then (25) can be wrote as an matrix operation

$$\mu = Q'^* S_1 . \tag{26}$$

The matrix $Q$ has the same structure and sizes as the matrix $D$ from the linear algebraic equation (13). But the matrix $D$ is very sparse and consists from zeros and units only, at the same time the matrix $Q$ is full filled.

## Differentiation of sinogram

In formula (25), which made the Radon transform, exist a numerical operation of differentiation of sinogram $S(p,\varphi)$ by detectors $p$. If $p$ considered as number of lines, and $\varphi$ — as number of columns, then numerical differentiation is more comfortable do by multiplication of $S(p,\varphi)$ from left on the corresponding matrix $M$, i.e. $S1(p,\varphi) = \partial S(p,\varphi)/\partial p = M\,S(p,\varphi)$.

As example below it is shown, how can be looked such matrix differential operator M, when it is a matrix with sizes 6×6.

From this example we can see that in the internal points of differentiated vector-column the derivative is approximated by a central differences, but in extreme points — one-side differences.

The matrices of differentiation operator M can be builded with any sizes. They are distinguished from Toeplitz matrices, whose have on diagonals the same values, by the first and last lines only.

$$M = \frac{1}{2h}\begin{pmatrix} -2 & 2 & 0 & 0 & 0 & 0 \\ -1 & 0 & 1 & 0 & 0 & 0 \\ 0 & -1 & 0 & 1 & 0 & 0 \\ 0 & 0 & -1 & 0 & 1 & 0 \\ 0 & 0 & 0 & -1 & 0 & 1 \\ 0 & 0 & 0 & 0 & -2 & 2 \end{pmatrix}$$

If $h$ is a step of discretization and matrix $M$ has size $N{\times}N$, then its generation is realized by commands:

```
c=zeros(N,1); c(2)=-1;
M=toeplitz(c,-c);
M(1,1)=-2; M(1,2)=2;
M(N,N)=2; M(N,N-1)=-2;
M=M/2/h;
```

## Code to testing the Radin transform

Code of function Qtom() to calculate by formula (24).

```
function Q=Qtom(np,nf,nx,ny)
Ap=90*pi/180; R=1.45; p=1:np; f=linspace(pi/2,3*pi/2,nf); alf=Ap*(1-2*(p-1)/(np-1))/2;
x=linspace(-1,1,nx); y=linspace(-1,1,ny); [ALF,F,X,Y]=ndgrid(alf,f,x,y);
Q=R*sin(ALF)+X.*sin(F-ALF)-Y.*cos(F-ALF); clear ALF F X Y; Q=Q./(Q.^2+0.008);
```

Scenario-file (testtom4) to test the Radon transform method. It makes execution by formula (25) and visualize the result.

```
np=60; nf=32; nx=40; ny=40; % testtom4
D=dtom3( np,nf,nx,ny);
Mu=zeros(nx,ny); Mu(8:12,8:12)=1; Mu(20:21,15:20)=0.5; Mu(27:37,18:30)= 0.3;
Mu(29:35,21:28)=0;
figure; imagesc(max(max(Mu))-Mu); colormap(gray);
```

```
Mu1=reshape(Mu,nx*ny,1); S=D*Mu1;
S=reshape(S,np,nf); figure; imagesc(max(max(S))-S); colormap(gray);
c=zeros(np,1); c(2)=-1; M=toeplitz(c,-c); M(1,1)=-2; M(1,2)=2;
M(np,np)=2; M(np,np-1)=-2; M=M/2; S1=M*S; clear D;
Q=Qtom(np,nf,nx,ny);
S1=reshape(S1,np*nf,1); Q=reshape(Q,np*nf, nx*ny);
Mu1=Q'*S1; Mu1=reshape(Mu1,nx,ny);
figure; imagesc(max(max(Mu1))-Mu1); colormap(gray);
figure; bar3(Mu1); colormap([1,1,1]);
```

Result of calculation by code testtom4

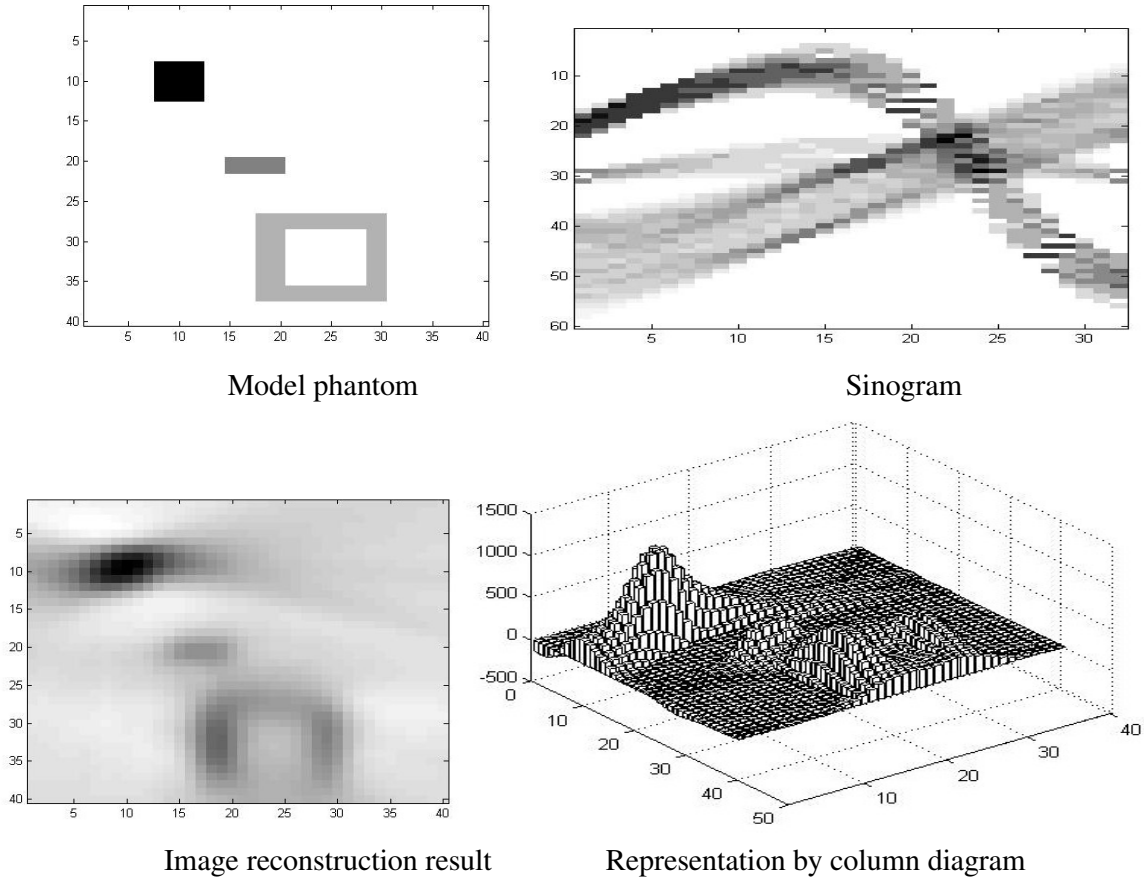

Model phantom          Sinogram

Image reconstruction result     Representation by column diagram

**Fig.13.**

We can't say that Radon transform by formula (25-26) is not working, but its reconstruction quality by direct algebraic method from equation (13) is ideal for the same model phantom.

## Verification of sensitivity of reconstructed image to the scanning parameters

The reason of good quality of image reconstruction by algebraic method is a usage of model phantom, when the sinogram and phantom artificially produced from the same matrix D. The pictures are not so optimistic when you are working with the real sinograms. Therefore we made the numerical experiment to analyze the sensitivity to the scanning parameters. Fig. 14 shows scheme of scanning from the radiation source placed in point C. The phantom region is marked by dashed contour, h is a half height of the detector array. The distances from phantom center to detector's array and to radiation source are designated by letters *a* and *d*, correspondingly. As unit distance we take one detector size. From other side, during coding the formulae describing the generation of matrix

D, we consider that the phantom's region is included in square -1<x,y<1. If the phantom is fixed, then radiation sources together with the detector array have to rotated around phantom's center on angle $\varphi_k$

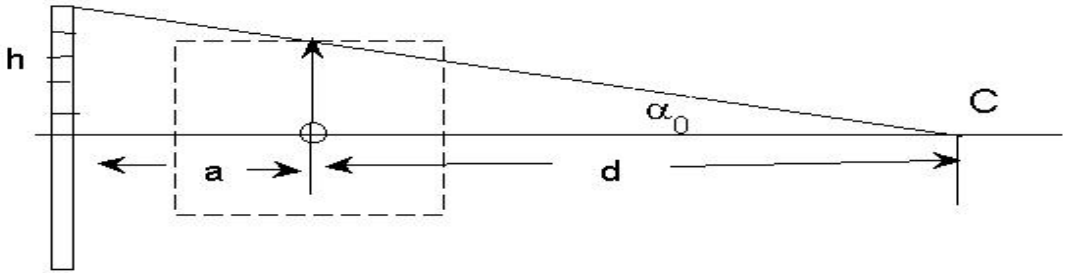

**Fig. 14**. Fan-beam scanning.

The scanning parameters in code dtom3 are a distance R (see Fig. 6) and aperture A, whose can be connected with values *a*, *d* and *h*. After that we get:

$$R = \frac{a+d}{h}, \qquad A = 2arctg\,\alpha_0 = 2arctg\,\frac{1}{R}. \qquad (27)$$

Below we show a fragment of code to calculate matrix D, where the distances *a* and *d* is included to the list of input parameters.

```
function D=dtom31(np,nf,nx,ny,a,d)
R=(a+d)*2/np; Ap=2*atan(1/R); f=linspace((2-1/16)*pi,0,nf);
x=linspace(-1,1,nx); y=linspace(-1,1,ny);
[F,X,Y]=ndgrid(f,x,y); D=angle(R-exp(-i*F).*(X+i*Y)); clear F X Y;
D=(D+Ap/2)/Ap; D=round(np*D);
D=reshape(D,1,nf*nx*ny); D=repmat(D,np,1)==repmat([1:np]',1,nf*nx*ny);
D=reshape(D,np*nf,nx*ny);D=sparse(D);
```

Let for the model phantom the sinogram was firstly generated with scanning parameters: *a*=300, *d*=500. Then, execution of scenario **testtom6** reconstructs the object for the different proposal about value of distance *d*.

```
np=160; nf=32; nx=40; ny=40; %testtom6;
D=dtom31( np,nf,nx,ny,300,500);
Mu=zeros(nx,ny); Mu(8:12,8:12)=1; Mu(20:21,15:20)=0.5; Mu(27:37,18:30)= 0.3;
Mu(29:35,21:28)=0;
Mu1=reshape(Mu,nx*ny,1); S=D*Mu1;

D=dtom31( np,nf,nx,ny,300,400);
Mu1=full(D.'*D)\(D.'*S); Mu1=reshape(Mu1,nx,ny);
subplot(2,2,1); imagesc(max(max(Mu1))-Mu1); colormap(gray);

D=dtom31( np,nf,nx,ny,300,490);
Mu1=full(D.'*D)\(D.'*S); Mu1=reshape(Mu1,nx,ny);
subplot(2,2,2); imagesc(max(max(Mu1))-Mu1); colormap(gray);

D=dtom31( np,nf,nx,ny,300,510);
Mu1=full(D.'*D)\(D.'*S); Mu1=reshape(Mu1,nx,ny);
subplot(2,2,3); imagesc(max(max(Mu1))-Mu1); colormap(gray);

D=dtom31( np,nf,nx,ny,300,600);
Mu1=full(D.'*D)\(D.'*S); Mu1=reshape(Mu1,nx,ny);
```

```
subplot(2,2,4); imagesc(max(max(Mu1))-Mu1); colormap(gray);
```

Fig. 15 shows the result of execution of the code **testtom6.**

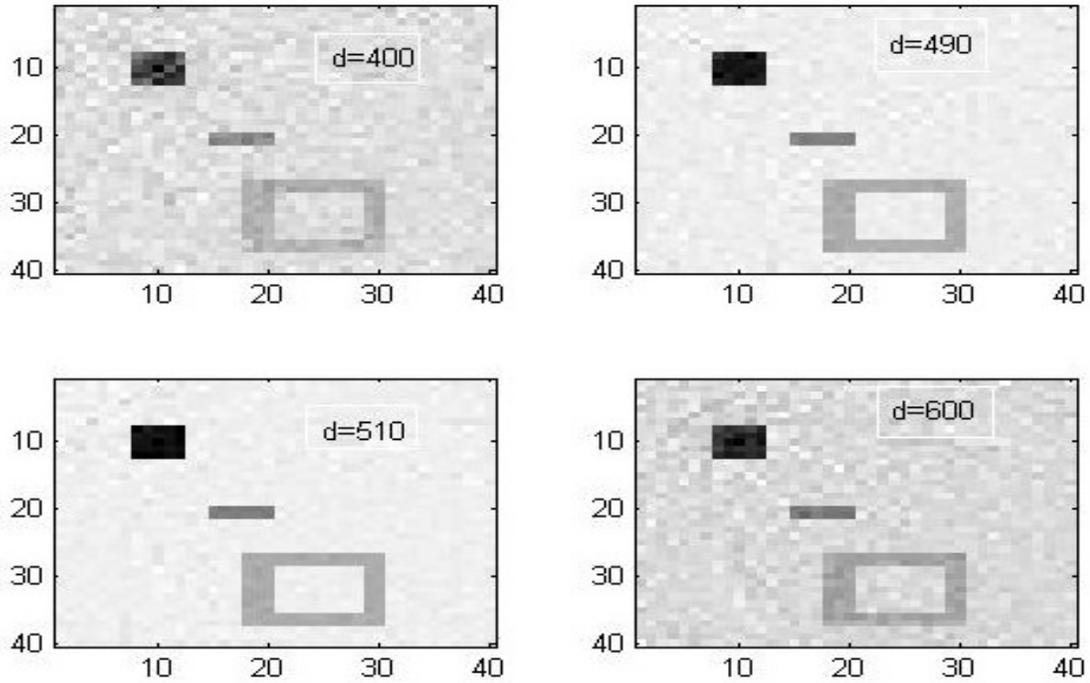

**Fig. 15.**

If the scanning parameters are known exactly (in this case *d*=500), then the algebraic method leads to ideal image reconstruction. From Fig. 15 we can see that for small deviations of distance *d* from the exact value the image reconstruction is enough satisfactory, but for large deviations there are appeared an artefacts.

**Reconstruction from the real sinogram**

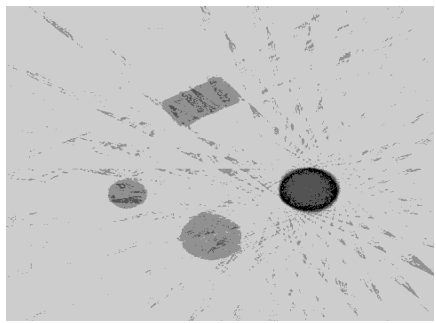

**Fig. 16.** Original phantom.

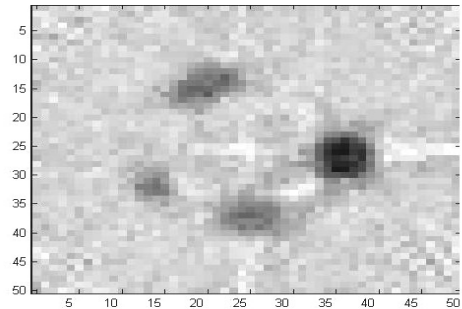

**Fig. 17.** Reconstruction result.

For the original phantom shown on Fig. 16, there was used the experimental sinogram. The scanning conditions were not known exactly. Though, the sonogram processing by algebraic algorithm testtom5 (with scanning parameters *a*=300, *d*=500) leads to similar result, shown on Fig. 17.

```
np=320; nf=32; nx=50; ny=50;%testtom5;
 ff=fopen('sinogram320x32.raw','r');  S=fread(ff,[320,32],'uint16'); fclose(ff);
 S=log(46000./S); D=dtom31( np,nf,nx,ny,300,500);   S=reshape(S,np*nf,1);
 Mu=(D'*D)\(D'*S); Mu=reshape(Mu,nx,ny);
imagesc(max(max(Mu))-Mu); colormap(gray);
```

## Conclusion

The main results of the report are following:

1. The mathematical approach of X-ray tomography task with usage matrix algorithm for image reconstruction was formulated.
2. In contrast to the known approaches, based on Radon transform usage, the matrix algorithm gives ideal reconstruction of the model phantoms. However, it is sensitive to the inaccuracy of scanning parameters. This drawback also inherent for other image reconstruction methods.
3. After further improvement those algorithms allow to solve a set of new practical problems.
4. We currently test the possibility of using the direct method of solving the inverse Radon equation in combination with the recently developed iterative reconstruction methods.